# Towards systematic near-threshold calculations in perturbative QFT

Fyodor V. Tkachov

Institute for Nuclear Research of Russian Academy of Sciences
60th October Ave. 7a, Moscow 117312 Russia

For any near-threshold asymptotic regime and for any Feynman diagram (involving loop and/or phase space integrals), a systematic prescription for explicitly constructing all-logs, all-powers (all-twists) expansions in perfectly factorized form with explicit integrals for coefficients, is presented. The distribution-theoretic nature of the method of asymptotic operation employed allows treatment of totally exclusive phase space in the same manner as loop integrals.

## Introduction        1

The purpose of this Letter is to present a summary of systematic recipe for construction of asymptotic expansions of Feynman diagrams near threshold values of kinematic variables. The recipe lays a foundation, heretofore lacking, for systematic higher-order calculations as well as for all-order construction of asymptotic expansions in operator/effective Lagrangian form. The range of applicability of the recipe includes many concrete applications such as near-threshold production of, say, electron-positron pair in QED (and many similar problems in the Standard Model); the small-$x$ problem in deeply inelastic scattering; etc.

The recipe is a culmination of the development of the technique of *asymptotic operation* (*As*-operation or simply AO) which has been the driving force behind the continuous progress in the systematic[1] studies of asymptotic expansions of Feynman diagrams in masses and momenta since before 1982 [1][2]. The Euclidean variant of AO yielded powerful calculational formulas for the short-distance expansion in the MS scheme [3; 4] and for mass expansions [5-8] that were used in a number of NNLO calculations in QCD and the Standard Model (e.g. [9-11]). The non-Euclidean extension of AO presented below is intended to play the same role with respect to the near-threshold problems as the Euclidean variant did with respect to those calculations.

Roughly speaking, Euclidean regimes correspond to cases when some masses and momenta are componentwise larger than others.[3] On the other hand, one deals with a truly non-Euclidean regime when some momenta have both large and small components, and the large components approach special values that result in a non-trivial overlap of singularities of different factors in the momentum-space integrand. Such situations correspond to thresholds from the point of view of analyticity properties of Feynman diagrams; cf. the discussion in [12].

The central difficulty of constructing asymptotic expansions of Feynman diagrams in masses and momenta is that formal Taylor expansions of integrands possess non-integrable singularities localized on variously intersecting manifolds in the space of integration momenta. The key observation (from which the entire theory of AO unfolds in a logical manner) is that the difficulty is a manifestation of the distribution-theoretic nature of the expansion problem [1][4], and that the crucial mathematical task is to find expansion of the integrand in the sense of distributions[5]. Asymptotic operation is a prescription that yields such an expansion for a given integrand (= product of propagators).

The structure of AO is fixed by the *extension principle* [1; 15] — a very general but essentially simple proposition related to the well-known Hahn-Banach theorem (see e.g. [17]). The resulting prescription is, roughly, as follows (see also Sec. 2 below):

(i) The formal expansion of the integrand should be supplemented with counterterms — linear combinations of (derivatives of) $\delta$-functions with coefficients that depend on the expansion parameter (cf. below Eq. 2.9).

(ii) Concrete integral expressions for the coefficients valid within the precision of expansion are obtained from the so-called consistency conditions [18; 15].

(iii) A fine-tuning of the consistency conditions to achieve purely power-and-log dependence of the coefficients of counterterms on the expansion parameter (the property of perfect factorization [3]).[6]

After the counterterms of AO are found, obtaining the corresponding all-order operator expansions is a matter of more or less straightforward combinatorics (cf. the Euclidean case [8]).

---

[1] "Systematic" here means all-logs, all-powers treatments complete with explicit calculational formulas in a form maximally suitable for practical applications.

[2] For a review and complete references to the original publications, as well as for comparison with the conventional methods such as the BPHZ theory and the technique of leading logarithmic approximation, see [2].

[3] To this class also belong degenerate cases with external large momenta are fixed on mass shell which simply corresponds to complex-valued large external momenta from the point of view of Euclidean space.

[4] This observation was influenced by Bogoliubov's analysis of the UV problem [13; 14].

[5] For precise definitions see [15; 16].

[6] Note that the deterministic logic of AO leads one step-by-step towards the solution. Contrast this with the BPHZ-type methods where one, in fact, needs to guess the result (the forest formula) as a whole in order to proceed to, say, formal proofs. For instance, the BPHZ-type interpretation of the general Euclidean expansions needed to borrow the explicit results that had been previously obtained within the framework of AO (for details see [19]).

Thus, the analytical focus of the entire theory of asymptotic expansions of Feynman diagrams is in finding the coefficients of the counterterms of AO in a form best suited for applications. The point is that, in general, asymptotic expansions are not unique (which is reflected in an arbitrariness of the expressions for coefficients of AO obtained from consistency conditions). Uniqueness is restored [15], however, if one imposes on expansions the requirement of *perfect factorization* [3] — which at the level of individual diagrams stipulates that the expansions run in pure powers and logarithms of the expansion parameter. Apart of its importance for applications (determination of power-suppressed corrections is impossible without it [20]), the uniqueness of power-and-log expansions has a number of technical advantages:

(i) AO commutes with multiplications by polynomials thus allowing one to ignore complications due to non-scalar particles;

(ii) expansions inherit all algebraic properties of the initial integrands (such as the Ward-Takahashi-Slavnov-Taylor identities);

(iii) maximal calculational simplifications; (iv) considerable advantages for a regularization-independent treatment [16; 21], which is potentially important for supersymmetric models.

The consistency conditions have the form of integrals involving some of the propagators of the original Feynman integrand as well as a test function whose behavior at the singular point is fixed but which is otherwise arbitrary (cf. Eq. 2.12). For the Euclidean case, it was found [18; 15] that it is sufficient to replace such test functions with suitable polynomials (to be understood as an appropriate limiting procedure; for a rigorous treatment see [16; 21]). It turned out that such a replacement modifies the expressions only within the precision of the expansion (which is always allowed), whereas the simple scaling of the integrand in momenta and masses ensures the power-and-log dependence of the resulting expression on the expansion parameter. This allowed us to include into systematic consideration the entire class of Euclidean asymptotic regimes [18] and subsequently to expand the scope of operator-expansion methods to such regimes.

As was emphasized from the very beginning [18; 15; 2], the scenario of AO is completely general and by no means limited to Euclidean cases. The specifics of the non-Euclidean regimes is that the singularities of integrands are localized on non-linear manifolds (light cones and mass shells) and that non-zero finite limiting values for external momenta break the usual scale invariance of integrands. As a consequence, the simple trick that yielded power-and-log dependence in the Euclidean case is no longer sufficient. However, it is important to understand that the consistency conditions are obtained from first principles without any restricting assumptions, and therefore possess all the flexibility to accommodate any additional requirement that one may lawfully impose — in particular, to perform an appropriate fine-tuning to achieve the required power-and-log dependence on the expansion parameter.

The resulting problem and its solution were identified in [2]: the problem consists in occurrence of the so-called *osculating singularities*[7], whereas the solution is given by the so-called *homogenization* — a secondary expansion that complements the consistency conditions for the coefficients of AO. The homogenization splits those coefficients into pieces with strict power-and-log dependence on the expansion parameter (powers needing not be integer). Below we expand the scenario of [2] by presenting explicit rules for the homogenization.[8]

It is remarkable that a self-contained analytical recipe for an individual diagram can be summarized in a rather compact universal form for arbitrary non-Euclidean regimes despite their larger analytical variety than in the Euclidean case. Such a description would not be possible without using the language of AO. However, there is both an increasing familiarity among physicists with the technique of AO, and a tendency to use it in various non-standard physical problems due to its power and flexibility (cf. [22-26])[9]. On the other hand, not all such works have been equally successful.[10] Therefore, it would be useful — prior to a more complete treatment which would require a substantially longer text — to give a summary of the procedure in a form suitable for calculations that can also serve as a starting point for derivation of all-order operator-form expansions for various regimes.

Before we proceed to formulas, a few remarks are in order.

(i)  One sometimes uses the term Minkowski space regimes (cf. [27]). However, Minkowski space per se allows both Euclidean regimes (cf. their treatment in Minkowski space in [12]), and a more complex class of non-Euclidean, or near-threshold regimes. I emphasize the distinction in order to avoid confusion due to vagueness of terminology and argumentation in many publications on the subject.

(ii) In some recent publications the term "threshold expansion" was misused to denote cases with kinematic parameters set exactly at threshold values with expansions running with respect to some other parameter (e.g. internal mass), or degenerate thresholds tractable by Euclidean methods. In this Letter we consider *true non-degenerate thresholds* (including non-zero ones) intractable by Euclidean methods, and expansions in a parameter that measures closeness to such a threshold.

(iii) The method of AO considers integrands in momentum space as distributions prior to integration. It is therefore ideally suited for studying physical problems where integration over the phase space of final state particles should not be performed in an explicit fashion. Indeed, the $\delta$-functions that describe the phase space — e.g. $\theta(p_0)\delta(p^2-m^2)$ etc. — are, from the distribution-theoretic point of view, equally acceptable factors alongside the standard causal propagators $(p^2-m^2+i0)^{-1}$. This opens a prospect for a systematic treatment of e.g. the problem of power corrections in jet-related shape observables in the context of precision measurements of $\overline{\alpha}_S$ according to the theoretical scenario outlined in [28] in analogy with the case of total cross section of $e^+e^- \to$ hadrons where the op-erator-product expansion can be used for that purpose [20]. Problems with exclusive phase space occur on a massive scale in physical applications.[11] The various cuts used for event selection are equivalent to various weights in phase space integrals, which means that the corresponding matrix elements squared are effectively treated as distributions in the momenta of the final state particles — a perfect setting for application of the distribution-theoretic technique of AO.

---

[7] i.e. singularities whose singular manifolds touch rather than intersect in a general fashion; for a detailed discussion see [12]. When a mixture of osculating and transverse intersections occurs, simple uniform scaling rules for description of the strength of the singularity no longer suffice.

[8] Ref.[2] outlined the worst-case scenario for non-Euclidean AO because a universal description of its structure was not yet available. Thus, the combinatorial complexity due to the homogenization seems now to be less severe — at least for some asymptotic regimes — than anticipated then.

[9] The integral version of AO [6; 8] (cf. also its first special case — the formulas for OPE coefficient functions in the MS scheme [3; 4]) is useful in situations where efficient automation is necessary, such as higher-order calculations etc.

[10] For instance, the description of AO in ref.[25] is incorrect whereas the authors of [26] independently found via AO a correct treatment of a concrete near-threshold problem.

## Description of the method 2

The prescriptions of AO are best described as a formal derivation rather than a final formula or theorem. The reason is that every step of the derivation has a simple concrete meaning enabling one to control correctness of formulas in each concrete situation, whereas blindly using a cumbersome final formula may result in gross errors.[12]

We follow the notations of [15; 16] and focus here only on the most difficult — analytical — aspect of the expansion problem; the diagrammatic interpretation depends on a concrete asymptotic regime and is a much simpler (combinatorial) issue anyway. We describe AO in the form with an intermediate regularization similarly to how the Euclidean case was treated in [15]. It is not difficult to convert the formulas into a regularization-independent form similar to [16] (further details specific to the non-Euclidean are given in [30]). In general, we have in view a combination of dimensional [31] and analytical [32] regularizations; the latter may be needed in the cases when the dimensional regularization alone is insufficient (cf. the example of [23]).

### The expression to be expanded 2.1

The collection of all integration momenta (loop and phase space) is denoted as $p$. The integrand of the diagram to be expanded in a small parameter $\kappa$ is represented as follows:

$$G(p) = \prod_{g \in G} g(p), \quad g(p) = \Delta_g(l_g(p,\kappa)), \qquad 2.2$$

where $\Delta_g(z) = (z \pm i0)^{-1}$ or $\delta(z)$ (each such factor will be referred to as "propagator"); $l_g(p,\kappa)$ is a second order polynomial of the momenta $p$ and masses (first order polynomials are allowed as a special case; cf. gauge boson propagators in non-covariant gauges). Note that we allow as factors causal propagators, their complex conjugates, and phase-space $\delta$-functions (for simplicity, the $\theta$-functions of phase space factors are omitted; to include them, it is sufficient to modify $\Delta_g$ accordingly in all formulas; cf. the example in Sec. 3.1). Various polynomials that may occur in the numerator (due to non-scalar particles and interactions with derivatives) are ignored because AO commutes with multiplication by polynomials [15]. The expression 2.2 comprises as special cases loop and unitarity diagrams.

*UV renormalization* is assumed to be performed in a massless scheme of the MS type and is treated following [33; 34; 21] as a subtraction from momentum space integra*nd* of its asymptotic terms (in the sense of distributions) for $p \to \infty$ (see [15; 16] for an exact interpretation of the large-$p$ limit involved). For practical purposes, it is sufficient to employ the MS scheme [35] (or any of the massless renormalization schemes), and treat unrenormalized UV-divergent integrals formally as convergent in the usual fashion (cf. the prescriptions of the Euclidean AO [15; 8]; a rigorous treatment of why this is possible is given in [16; 21]).

### Expansion parameter and asymptotic regimes 2.3

The expression 2.2 depends on external parameters such as masses, momenta of incoming particles etc. It is assumed that some of the momentum components and/or masses are small compared to others. The small parameter (one with respect to which the expansion is to be performed) will be denoted as $\kappa$. In general, one assumes that some of the external parameters — masses or momenta — tend to specific values (zero or not), and that the differences between the external parameters and their limiting values are of order $\kappa$ (extension to cases with several scales of the form $\kappa^n$ is straightforward). The limiting values of external momenta need not be zero componentwise. Some examples of the constructs mentioned in the description given below, are presented in Sec. 3 (for further examples see [23] and [36]).

Our prescriptions are valid irrespective of what kind of threshold the chosen asymptotic regime corresponds to — perhaps, none at all in which case there will simply be no singularities requiring addition of non-trivial counterterms.

Note also the following rule for the problems with explicit phase space: If some phase space momentum components are to be treated as small, i.e. $O(\kappa)$, they should be made $O(1)$ by appropriate rescaling before applying the procedures of AO.

### Formal expansion 2.4

The construction of AO begins with the formal (usually, but not necessarily, Taylor) expansion of the integrand in powers of $\kappa$: $G(p) \to \mathsf{T}_\kappa \circ G(p)$. Each factor is expanded separately, the results are formally multiplied and reordered in increasing powers of $\kappa$. The terms of the resulting series possess, in general, non-integrable singularities which have to be examined in the geometrical and analytical aspects.

### Geometric classification of singularities of the formal expansion 2.5

Each formal expansion $\mathsf{T}_\kappa \circ g(p)$ of each factor $g(p)$ of the initial product 2.2 is singular on the manifold $\pi_g$ described by $l_g(p,0) = 0$. The aggregate singular manifold of $\mathsf{T}_\kappa \circ G(p)$ is $\cup_{g \in G} \pi_g$. In general, the latter is singular in the sense of differential geometry, so one splits it into non-singular components $\pi_\gamma$ (labeled by an index $\gamma$; each such component is a smooth open manifold). To each component there corresponds a subproduct $F_\gamma(p) \subset G(p)$ (but unlike the Euclidean case, here different $\pi_\gamma$ may correspond to the same subproduct; e.g. the apex of the light cone and its two cones correspond to the same propagator). $F_\gamma(p) \equiv \gamma(p)$ contains all the factors from $G$ that are singular everywhere on $\pi_\gamma$. Denote as $G \setminus \gamma(p)$ the product of all factors from $G$ that do not belong to $\gamma$. One may say that $\gamma$ represents a "subgraph"; its corresponding product of propagators $F_\gamma(p) \equiv \gamma(p)$ is determined uniquely.[13]

---

[11] I am indebted to S. Jadach for explaining to me this point.

[12] Incidentally, Collins et al. [25] attempted to describe a non-Euclidean example of AO. The result is a bizarre text which makes very little sense beyond vaguely echoing the discussions I had with John Collins during my three visits to Penn State (I believed we were discussing application of the technique of AO to the Sudakov problem; cf. [23]). Several of the formulas of [25] (together with the accompanying textual descriptions) are simply incorrect. This calls for a critical reexamination of the "proofs" of QCD factorization theorems, which I intend to do elsewhere [29].

### Analytical structure of singularities                          2.6

Here we have to set rules for power counting and — simultaneously — to define the so-called homogenization — a secondary expansion needed in non-Euclidean situations to reduce the coefficients of counterterms of AO to power-and-log form. The discussion below is in the context of a given subgraph $\gamma$. One considers a general point $p_0$ of $\pi_\gamma$ and the behavior of $F_\gamma(p)$ when $p \to p_0$ along directions that are transverse to $\pi_\gamma$. The components of $p$ that are tangential to $\pi_\gamma$ are "spectators" and are to be ignored. After introducing appropriate local coordinates near $p_0$ and redefinitions, we may assume in what follows that $p$ does not have spectator components and that $\pi_\gamma = \{0\}$.

*Scaling*   Scale $p_i \to \lambda^{n_i} p_i$, $\kappa \to \lambda^{n_\kappa}\kappa$ so that:

(i) all scaling exponents $n_\alpha$ are positive integer;

(ii) for each $g \in \gamma$, $l_g(p,\kappa)$ scales as $\lambda^{n_g}\left[l_g^{\mathrm{main}}(p,\kappa) + O(\lambda)\right]$, and the scaling exponent for $\kappa$ is the minimal value ensuring this for given $n_i$;

(iii) if $\pi_g^{\mathrm{main}}$ is the manifold on which $l_g^{\mathrm{main}}(p,0) = 0$, then $\pi_\gamma^{\mathrm{main}}$ — defined as $\cap_{g\in\gamma}\pi_g^{\mathrm{main}}$ i.e. the set of all $p$ such that $l_g^{\mathrm{main}}(p,0) = 0$ for all $g \in \gamma$ — coincides with $\pi_\gamma = \{0\}$. The latter means, in particular, that the collection of all $l_g^{\mathrm{main}}(p,\kappa)$, $g \in \gamma$ depends on all components of $p$.

These properties ensure that each step of our expansion procedure is a mathematically correct transformation. A remarkable fact is that the scaling satisfying (i)–(iii) need not be unique (e.g. in the case of radiative corrections to the example of Sec. 3.1). The beautiful mathematical mechanism of how different scalings result in the same final answer, is mentioned below after the definition of homogenization — we have already had opportunities [15] to emphasize a remarkable stability of the method of AO that yields correct results even in counterintuitive cases as long as one applies it in a systematic manner. Different correct scalings do differ in the number of intermediate steps leading to the (same) final result. An optimal definition is as follows: Split $l_g(p,0) \to L'_g(p) + L''_g(p)$ with $L'_g(p) = O(p)$, $L''_g(p) = O(p^2)$. Then split $p \to (X,Y,Z)$ so that all $L'_g(p)$ depend on, and only on $X$ and all $l_g(p,0)$ with $L'_g(p) \equiv 0$ are independent of $Z$. Then $X$ and $Y$ scale with $\lambda^2$, and $Z$ scales with $\lambda$.[14] As a simple check, this rule correctly yields a uniform scaling in all components of $p$ and $\kappa$ both when all $l_g(p,\kappa)$ are linear functions, and when they are all purely quadratic functions (the Euclidean case).

*Power counting*[15]   One performs the power counting to determine the strength of singularity in each term of $\mathbf{T}_\kappa \circ G(p)$ near generic points of each $\pi_\gamma$. For that, one drops from denominators (and/or arguments of $\delta$-functions) all but those components of $l_g(p,0)$ that scale with the lowest power of $\lambda$, and introduces into the numerator the factor $\lambda^{2\dim X + 2\dim Y + \dim Z}$ that corresponds to the scaling of the integration measure. Collecting all powers of $\lambda$ one obtains an overall factor $\lambda^{-\omega_\gamma}$ where $\omega_\gamma$ can be appropriately called the singularity index of the subgraph (e.g. $\omega_\gamma = 2$ corresponds to quadratic divergence etc.).

### Homogenization                                                 2.7

The *homogenization parameter* $\xi_\gamma$ is introduced as follows: (i) scale $l_g(p,\kappa)$, $g \in \gamma$ as described; (ii) drop the overall $\lambda^{n_g}$; (iii) replace $\lambda$ with $\xi_\gamma$. The operation of *homogenization* (denoted as $\mathbf{H}_\gamma$) is as follows: (i) introduce $\xi_\gamma$ as just described; (ii) expand in $\xi_\gamma$; (iii) set $\xi_\gamma \to 1$ in the result. $\mathbf{H}_\gamma$ is meant to be applied to integrals similar to those we set out to expand from the very beginning (see below Eq. 2.12). Therefore, the expansion in $\mathbf{H}_\gamma$ must in general be performed in the sense of distributions — requiring the entire machinery of AO with another level of homogenization etc. Since at each step the expansion problem simplifies (dimensionality of the integration space is reduced), the recursion stops correctly after a finite number of levels of homogenization. In many interesting cases, however, non-trivial singularities requiring counterterms do not occur and this expansion degenerates into a simple Taylor expansion.

The occurrence of secondary expansions with their corresponding homogenizations is behind the mechanism that ensures independence of final results of the choice of scaling (provided the latter satisfies the three conditions mentioned in Sec. 2.6): non-trivial counterterms for secondary homogenizations yield expressions that correspond to alternative scalings, whereas the "formal" part of the homogenization expansion yields zero by explicit integration. Demonstrating this mechanism in detail requires a much more detailed exposition than we can afford here.

---

[13] The pinch/non-pinch classification of singularities (the usual starting point of the conventional analyses; cf. e.g. [27]) is actually irrelevant for the analytical study of the singularities in general, and for the construction of AO in particular: the non-pinched singularities simply correspond to parts of singular manifolds where the corresponding counterterms nullify (which can be deduced e.g. directly from their expressions). Such a nullification is rather accidental from the point of view of analytical structure of singularities.

[14] For general non-quadratic (cubic etc.) functions one may recur to a technique based on Newton's polyhedra similar to that used in the theory of singularities of differential mappings [37].

[15] This is regarded as a technical problem in the context of the theory of QCD factorization theorems [27], esp. in the case of non-leading power corrections and mixed soft/collinear singularities. Our rules should settle the issue (see also [30]).

### Structure of asymptotic operation 2.8

Recall the general formula for AO [15]:

$$\mathsf{As}_\kappa \circ G(p) = \mathsf{T}_\kappa \circ G(p) + \sum_\gamma \mathsf{E}_\gamma(p) \times \left[ \mathsf{T}_\kappa \circ G \setminus \gamma(p) \right]. \quad 2.9$$

It is rather natural: each singularity of the formal expansion $\mathsf{T}_\kappa \circ G(p)$ receives a counterterm $\mathsf{E}_\gamma(p)$ localized on the corresponding singular manifold $\pi_\gamma$. So, the summation runs over all subgraphs $\gamma$, and $\mathsf{E}_\gamma(p)$ have the form

$$\mathsf{E}_\gamma(p) = \sum_\alpha E_{\gamma,\alpha}(\kappa) \times \delta_{\gamma,\alpha}(p), \quad 2.10$$

where summation runs over a complete set of $\delta$-functions localized on $\pi_\gamma$, and $E_{\gamma,\alpha}(\kappa)$ in general are — unlike the Euclidean case where each $E_{\gamma,\alpha}(\kappa)$ is proportional to one integer power of $\kappa$ — series in (non-integer) powers of $\kappa$ with coefficients that are polynomials of $\log \kappa$. As was already emphasized, finding those coefficients is the central analytical task of the theory of asymptotic expansions of Feynman diagrams.

### Consistency conditions for the coefficients 2.11

The finding of $E_{\gamma,\alpha}(\kappa)$ is performed according to the same general recipe as in [15], i.e. via consistency conditions. Suppose one wishes to construct the coefficients for $E_{\Gamma,\alpha}(\kappa)$ for one subgraph $\Gamma$. One assumes that for all $\gamma < \Gamma$ the construction has been performed (the usual ordering of subgraphs with respect to increasing codimensionality of singular manifolds is assumed here; cf. [15]). An appropriate choice of coordinates ensures that the singularity one is after is localized at $p = 0$. Then the coefficients are given by the following formulas (cf. sec. 12.3 in [15] and eq. (20.5) in [16]):

$$\begin{aligned} E_{\Gamma,\alpha}(\kappa) &= \lim_{\Lambda \to \infty} \int dX dY dZ\, \Phi(X\Lambda^{-2}, Y\Lambda^{-2}, Z\Lambda^{-1}) \\ &\qquad \times \mathcal{P}_\alpha(p)\, \mathsf{H}_\xi \circ \left[ 1 - \mathsf{As}'_\kappa \right] \circ \Gamma \\ &= \int dp\, \mathcal{P}_\alpha(p)\, \mathsf{H}_\xi \circ \Gamma(p,\kappa). \end{aligned} \quad 2.12$$

(The polynomials $\mathcal{P}_\alpha$ form a complete dual set for the $\delta$-functions $\delta_{\Gamma,\alpha}$ — exactly as in the Euclidean case. $\mathsf{As}'_\kappa \circ \Gamma$ differs from $\mathsf{As}_\kappa \circ \Gamma$ by absence of the term with $\gamma = \Gamma$ in the corresponding sum 2.9.) The first expression demonstrates how the intermediate cutoff is removed. Notice the asymmetry of the cutoff which corresponds to the asymmetry of the scaling. The role of $\mathsf{H}_\xi$ (a new element compared to the Euclidean case) is to split the coefficient into terms with pure power-and-log dependence on $\kappa$ (which motivates its definition). The second expression takes into account the nullification of the subtracted ("shadow") terms in dimensional/analytic regularization, which is due to their pure-power behavior under the (asymmetric) scaling.[16,17] The power of $\kappa$ for each term in the last expression in 2.12 is determined by scaling out $\kappa$ using the scaling rules for $\Gamma$ already explained.

An important point to remember is that the construction of asymptotic expansions (including AO) is always, strictly speaking, carried out for a particular finite precision $O(\kappa^N)$. This implies that one subtracts only the series to that precision in the integrand of 2.12. Correspondingly, the counterterm is also defined at that step only within that precision, whereas expansion implied by $\mathsf{H}_\xi$ — if carried too far — would generate terms of excessive precision $O(\kappa^{N+n})$ — terms that would also be divergent in the UV region! One can verify by power counting, however, that all the contributions of precision $O(\kappa^N)$ are exactly those whose (formal) UV convergence is ensured by the subtraction. In the end one can forget about the restriction $O(\kappa^N)$ because the formula (the last expression in 2.12) is independent of it.[18]

Another important point concerns the diagrammatic interpretation of the subtraction in the first expression in 2.12. In the Euclidean case, it was shown [34; 21] that the subtractions of this sort exactly correspond to the standard Bogoliubov UV $R$-operation — in agreement with the fact that the integrals that occurred there for $E_{\Gamma,\alpha}$ were exactly diagrams with local operator insertions. In the general non-Euclidean situation, there is no universal (i.e. valid for all regimes) operator characterization for the integrals that occur after the homogenization (the expansion $\mathsf{H}_\xi$ distorts standard propagators in different ways for different asymptotic regimes). But the prescription for their UV renormalization *is always determined uniquely by the structure of subtractions in 2.12*.

As a last remark, an interesting technical point may be mentioned. Namely, despite the presence of the additional expansion $\mathsf{H}_\xi$ in Eq. 2.12 compared to the Euclidean case, when one integrates out the $\delta$-functions similarly to the procedure of sec. 5.3 of [8] to obtain the AO in integral form, this additional series blends into a single Taylor expansion with the expansion resulting from derivatives of $\delta$-functions (cf. eq.(5.8) in [8]). This is easily explained if one notices that the non-Euclidean prescriptions remain valid in the general non-linear/non-uniform case, so that the "spectator" product $G \setminus \Gamma$ can be formally absorbed into $\Gamma$ and treated as a "deformation" of the latter, and when obtaining the AO in integral form, only those counterterms of the AO would survive that are proportional to $\delta$-functions without derivatives. But then the expansion $\mathsf{H}_\xi$ would be the *only* expansion remaining in the end! (Note, however, that the expansion implied by $\mathsf{H}_\xi$ affects the entire $G$ — in contrast with the Euclidean case where only certain subgraphs are thus expanded; cf. the expansions of "heavy knots" in [8].) An amusing implication is that use of the general non-Euclidean results would actually simplify the construction of Euclidean AO in integral form.

---

[16] Nullification of integrals of the form $\int d^D p\, p^{-2\alpha} = 0$ is well-known. It is due to the fact that the dimensional regularization preserves formal scalings. In the new situation, we are dealing with zero integrals of, very roughly, the form $\int d^{D-2} p_\perp dp_+ \left( A p_+ + p_\perp^2 \right)^{-2\alpha} = 0$. The underlying reason for their nullification is the same — preservation of scaling properties by the dimensional/analytical regularization, although now the scaling is non-uniform in different components of $p$.

[17] To obtain the corresponding regularization-independent formula one replaces $\mathsf{As}'_\kappa \to \tilde{\mathsf{r}}_f \circ \mathsf{As}'_\kappa$ where $\tilde{\mathsf{r}}_f$ is an appropriate generalization of the corresponding Euclidean subtraction operator; see [16] and [30].

[18] However, a great caution must be exercised when establishing correspondence between the terms of the last expression in 2.12 and the singularities of the formal expansion $\mathsf{T}_\kappa \circ G$. The correspondence is rather tricky which fact caused some confusion in the literature (see a discussion in [19]).

## Examples 3

The two examples we consider are related to well-known simple integrals so that checks are possible, yet involve explicit phase space so that a systematic treatment via any other method would be problematic. In both examples, obtaining higher terms of the expansion is tedious but entirely straightforward — an exercise that is left to an interested reader. Substantially more involved examples will be presented elsewhere [38; 39].

### Threshold $Q^2 \sim 4m^2$ for the kinematics $Q \to m+m$    3.1

This example corresponds to a near-threshold creation of a pair (e.g. $e^+e^-$ by a photon). Remember that asymptotic operation commutes with multiplication by polynomials so that non-scalar numerators are ignored in our example.[19] The phase space is represented as

$$\int d^D p \, w(p) \, \delta_+\!\left(p^2 - m^2\right) \delta_+\!\left((Q-p)^2 - m^2\right), \qquad 3.2$$

where $\delta_+(k^2 - m^2) = \theta(k_0)\delta(k^2 - m^2)$, $Q$ is the momentum of the "photon" that decays into the pair, and the arbitrary weight $w$ (corresponding to arbitrary cuts experimentalists may use for event selection) means that the phase space is treated as totally exclusive, so that one essentially deals with the product of $\theta$'s and $\delta$'s interpreted as a distribution. We are going to extract the first non-trivial (square root) contribution using the prescriptions set forth above.

To describe the asymptotic regime we choose $Q = (2m + \kappa, \mathbf{0})$ with $\kappa \to 0$. Then

$$\kappa = (Q^2 - 4m^2)/4m - (Q^2 - 4m^2)^2/(4m)^3 + \ldots \qquad 3.3$$

It is convenient to perform a shift $p \to p + \tilde{m}$, $\tilde{m} = (m, \mathbf{0})$. The formal expansion of the product in $\kappa$ then is:

$$\delta_+\!\left((p+\tilde{m})^2 - m^2\right) \delta_+\!\left((\tilde{m}+\tilde{\kappa} - p)^2 - m^2\right)$$
$$= \delta_+\!\left(p_0^2 + 2mp_0 - \mathbf{p}^2\right) \delta_+\!\left(p_0^2 - 2mp_0 - \mathbf{p}^2\right)$$
$$+ 2\kappa[-p_0 + m]\delta_+\!\left(p_0^2 + 2mp_0 - \mathbf{p}^2\right)\delta'_+\!\left(p_0^2 - 2mp_0 - \mathbf{p}^2\right) + \ldots \quad 3.4$$

The singularity at the point $p = 0$ is seen from the fact that the arguments of the two $\delta$-functions reduce to the form $\delta(p_0)\delta(\mathbf{p}^2)$, with the second $\delta$ ill-defined in the case of higher derivatives (the singularity is regulated by the dimensional regularization). To do the power counting according to the prescriptions of Sec. 2, one identifies $X \leftrightarrow (p_0, \kappa)$, $Y \leftrightarrow \mathbf{p}$ with $Z$ empty. The first product of $\delta$'s is convergent whereas the second one is linearly divergent. In any case, a counterterm of the form $E(\kappa)\delta(p)$ is necessary. Evaluating $E(\kappa)$ according to the given recipe (the homogenization consists simply in Taylor-expanding with respect to the quadratic terms in $p_0^2$ etc.), we find:

$$E(\kappa) = \int d^D p \, \delta\!\left(2mp_0 - \mathbf{p}^2\right)\delta\!\left(2m(\kappa - p_0) - \mathbf{p}^2\right)$$
$$= \tfrac{1}{4m} \times \tfrac{\pi^{3/2-\varepsilon}}{\Gamma(3/2-\varepsilon)} \times (m\kappa)^{1/2-\varepsilon}, \qquad 3.5$$

where $\varepsilon = \tfrac{1}{2}(4 - D)$. The final result for 3.2 is as follows:

$$\int d^D p \, w(p) \, \delta_+\!\left(p^2 - m^2\right)\delta_+\!\left((Q-p)^2 - m^2\right)$$
$$= \tfrac{\pi}{4m}\sqrt{Q^2 - 4m^2}\, w(\tilde{m}) + O\!\left(Q^2 - 4m^2\right). \qquad 3.6$$

(The integral of the first term of the formal expansion — the first term on the r.h.s. of 3.4 — is zero for smooth $w$.) This result is checked by noticing that the case $w = 1$ corresponds to the imaginary part of a simple self-energy diagram; cf. the explicit result in [40], eq.(24.5).

### The behavior of $2 \to 2$ at $s \to +\infty$    3.7

Our second example corresponds to the matrix element squared with exclusive phase space for $k_+ + k_- \to k'_+ + k'_-$ (where $k'_\pm = k_\pm \pm p$ ) via simplest $t$-channel exchange of a scalar particle with momentum $p$ at large $s = (k_+ + k_-)^2$. The example is motivated by the large-$s$/small-$x$ problem in QCD [41], so all particles (partons) are massless and the external partons are slightly off-shell (time-like). The expression to be expanded is:

$$s \int dp \, w(p) \, \delta_+\!\left((k_+ + p)^2\right)\delta_+\!\left((k_- - p)^2\right)p^{-2}, \qquad 3.8$$

where $\delta_+\!\left((k_+ + p)^2\right) = \theta(k_+^0 + p^0)\delta\!\left((k_+ + p)^2\right)$ etc., and we have replaced for simplicity the exchange particle's squared propagator $p^{-4}$ by $p^{-2}$ (imagine e.g. that there occurred a cancellation with $p^2$ from a non-scalar numerator). $w$ is an arbitrary weight that describes cuts, observables etc. The asymptotic regime is described by $s \to \infty$ with $k_\pm^2, p^2 = O(1)$. Introduce two light-like vectors $q_\pm$ such that $2q_+ q_- = 1$ and $k_\pm = \sqrt{s}\, q_\pm + O(s^{-1/2})$. Taking out $\sqrt{s}$ from each $\delta_+$ and denoting $\kappa = s^{-1/2}$, rewrite 3.8 as

$$\int dp \, w(p) \, p^{-2} \, \delta_+\!\left(2q_+ p + \kappa k_+^2 + \kappa p^2 + O(\kappa^2 p)\right)$$
$$\delta_+\!\left(-2q_- p + \kappa k_-^2 + \kappa p^2 + O(\kappa^2 p)\right). \qquad 3.9$$

The first term of the formal expansion is:

$$\int dp \, w(p) \, \delta_+\!\left(2q_+ p\right)\delta_+\!\left(-2q_- p\right)p^{-2} + O(\kappa). \qquad 3.10$$

Singularities are localized on $\pi_\pm = \{p = zq_\pm; z \in \mathbb{R}\}$ and $\pi_0 = \{0\}$. The two collinear singularities $\pi_\pm$ are similar to those considered in [23]. Following the above rules, one finds the appropriate scaling (in the Sudakov coordinates): $p_\mp \to \lambda^2 p_\mp$, $p_\perp \to \lambda p_\perp$, $\kappa \to \lambda^2 \kappa$ (for $\pi_\pm$ the component $p_\mp$ is spectator and does not scale). One finds that in the leading order the two singularities are logarithmic, so only $\delta$-functions without derivatives have to be added to 3.10:

$$\int_{-\infty}^{+\infty} dz \, c_+(z,\kappa)\delta(p - zq_+) + \int_{-\infty}^{+\infty} dz \, c_-(z,\kappa)\delta(p - zq_-). \qquad 3.11$$

---

[19] Of course, they do play a role in determining which particular terms from the asymptotic expansion contribute to a particular process at a particular precision with respect to the small parameter. But our aim here is only to demonstrate the essential analytical mechanism.

The corresponding coefficients are easily found:

$$c_\pm(z,\kappa) = \int d^D p\, \delta_+(\pm 2pq_\pm + \kappa k_\pm^2)\, \delta(z - 2pq_\mp)\, p^{-2}$$
$$\sim \frac{1}{\varepsilon}(\mp z)_+^{-\varepsilon}\left(\kappa k_\pm^2\right)^{-\varepsilon}, \qquad 3.12$$

where $A_+^\alpha = \theta(A > 0) A^\alpha$ and we have dropped irrelevant coefficients. However, when one multiplies the collinear counterterms by the corresponding $G \setminus \gamma_\mp = \delta_+(\mp 2q_\mp p)$, one finds that the product vanishes because $c_\pm(z) \sim z^{-\varepsilon}$ (cf. the case with a non-zero mass where a similar product was ill-defined [23]).

The scaling for the singularity localized at $\pi_0 = \{0\}$ is uniform in all components $p \to \lambda p$, $\kappa \to \lambda \kappa$. The counterterm then is $c_0(\kappa)\delta(p)$ with

$$c_0(\kappa) = \int d^D p\, \delta_+\!\left(2q_+ p + \kappa k_+^2\right) \delta_+\!\left(-2q_- p + \kappa k_-^2\right) p^{-2}$$
$$\sim \frac{1}{\varepsilon} \kappa^{-2\varepsilon}. \qquad 3.13$$

Due to all the $\delta$-functions, the integrals in this example are performed very easily.

Finally, the expansion is:

Eq. 3.8 = Eq. 3.10 + $c_0(\kappa) w(0)$. \qquad 3.14

Performing integration in 3.10 with $w = 1$ one finds that the poles $\varepsilon^{-1}$ in 3.14 cancel (as expected). One sees that the $\log \kappa$ contribution is associated with the diffractive region $p \sim 0$.

If one had normal non-cut propagators instead of the two $\delta$-functions in 3.8, the vanishing of collinear counterterms would not occur, which would give rise to the well-known $\log^2 \kappa$ terms (taking imaginary part eliminates one logarithm). Otherwise the integrals are just a bit more involved, and the correctness of calculations is easily checked by comparison with the well-known analytical results (see e.g. [42]).

For Monte-Carlo type applications one needs to recast the results into a regularization-independent form. The corresponding prescriptions are derivative from the power-counting rules of Sec. 2.6 and follow the general pattern of [16; 21]. More details will be given elsewhere [30].

## Conclusions 4

The recipe for the non-Euclidean (near-threshold) asymptotic operation given above is a very general one, is valid practically for any non-Euclidean asymptotic regime, and for any individual Feynman diagram, including unitarity diagrams with cut propagators as well as diagrams in non-covariant gauges, heavy-quark and non-relativistic effective theories, etc. But despite its generality, it relies on few analytical principles, which is important for calculationists who wish to have a complete intellectual control over what they are doing (cf. e.g. [22; 24; 26]). In many applications (namely, automated higher-order calculations and derivations of all-order near-threshold expansions in operator/effective Lagrangian form) one would need the rules for AO in integral form similar to the Euclidean ones found in [5-8], that are currently in wide use (cf. [9-11]). Such rules are not difficult to obtain — but then the universality of formulas is lost and each regime has to be treated separately. Studying the operator form of expansions for various non-Euclidean (near-threshold) regimes and performing the corresponding calculations will be a topic of active research in the coming years [26; 38; 39; 43].

*Acknowledgements* I am indebted to V.I.Borodulin, G.Jikia and N.A.Sveshnikov for an extremely important support. I thank D.Broadhurst and P.Nason for pointing out the importance of studying non-zero thresholds in QED and the Standard Model, S.Jadach for emphasizing the importance of exclusive treatment of phase space, A.V.Radyushkin for bibliographic advice, and F.Berends for an encouraging laugh. This work was supported in part by the Russian Foundation for Basic Research under grant 95-02-05794.